\documentstyle[epsfig,float]{aipproc}

\begin{document}

\title{Magnetic Field Limits on SGRs}

\author{R.E. Rothschild $^1$, D. Marsden$^{1,2}$, \& R.E. Lingenfelter
$^1$}

\address{1) Center for Astrophysics and Space Sciences, University of
California, San Diego, La Jolla, CA USA,\\ 2) presently NAS/NRC
Research Associate at Goddard Space Flight Center}

\maketitle

\begin{abstract}

We measure the period and spin-down rate for SGR 1900+14 during the
quiescient period two years before the recent interval of renewed burst
activity. We find that the spin-down rate doubled during the burst
activity which is inconsistent with both mangetic dipole driven spin 
down and a magnetic field energy source for the bursts.  We also
show that SGRs 1900+14 and 1806-20 have braking indices of $\sim$1 
which indicate that the spin-down is due to wind torques and not magnetic
dipole radiation.  We further show that a combination of dipole radiation,
and wind luminosity, coupled with estimated ages and present spin
parameters, imply that the magnetic fields of SGRs 1900+14 and 1806-20
are less than the critical field of 4$\times$10$^{13}$ G and that the
efficiency for conversion of wind luminosity to x-ray luminosity is
$<$2\%.

\end{abstract}

\section{ Spin-Down History of SGR 1900+14}

The spin-down of SGR 1900+14 from 1966 September to 1999 April (Figure
1) is characterized by three intervals of time for which the spin-down
rate was essentially constant within the interval.  This
characterization of the SGR 1900+14 spin-down is based upon both
direct measurements of $\dot{P}$ as part of the period determination,
and upon differences in measured spin periods between two different
observations.  The first interval begins with the RXTE observation in
September of 1996 and ends with the ASCA observation at the beginning
of May, 1998.  The mean spin-down $\dot{P} \sim$ 6$\times$10$^{-11}$
s/s \cite{mar99a,hur99,woo99a}.  The second interval begins with the
onset of bursting on May 26, 1998 and continues until mid-September
1998.  The mean spin-down during this time $\dot{P}
\sim$13$\times$10$^{-11}$ s/s \cite{kou99,mar99a,mur99}.  The third
interval begins in mid-September 1998 and continues at least until
March 30, 1999.  The mean spin-down rate at that time is again $\sim$
6$\times$10$^{-11}$ s/s \cite{woo99b}.

Woods et al. \cite{woo99b} (1999b) have suggested that the data may be
consistent with a discontinuous spin-down event during the second
interval as a result of the giant burst of August 27, 1998, as opposed 
to a doubling of $\dot{P}$ during the entire second interval \cite{mar99a}.  
This suggestion, however, appears to be at odds with  the measurement of
$\dot{P}$=(11.0$\pm$1.7)$\times$10$^{-11}$ s/s in early June, 1998
\cite{kou99}, approximately 3 months before the Superburst and with the
RXTE/ASCA determination of $\dot{P} \sim$ 10$\times$10$^{-11}$ s/s just
after the event \cite{mur99}. A $\dot{P}$ variation of more than a 
factor of 3 was also indicated in the timing measurements of SGR 1806-20
\cite{kou98}.

\begin{figure} 
\centerline{\psfig{file=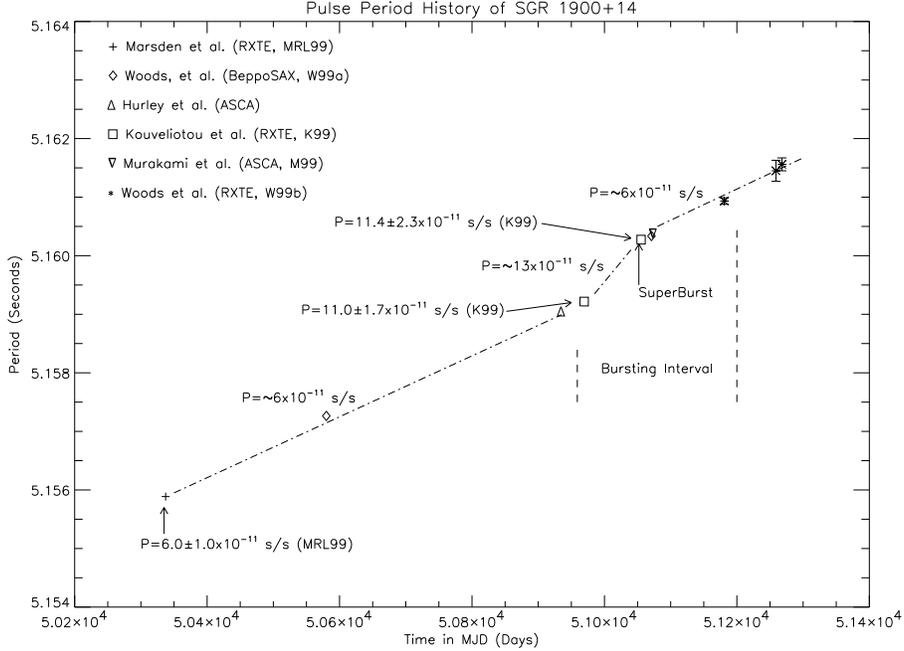,
angle=90, width=12cm}} 
\caption[]{The pulse period history for SGR
1900+14 versus time. All of the published values of the pulse period
are given along with the three measurements of $\dot{P}$ made as part
of the period determination analysis. The mean spin-down in the three
time intervals are also given.} 
\end{figure}

Very similar variations in the spin down rate $\dot{P}$ have been
observed in AXPs 1E1048.1-5937 and 1E2259+58 (see review by Stella,
Israel, and Mereghetti \cite{sim98}). Such behavior argues strongly
against SGR spin-down by magnetic dipole radiation, since that would
require large increases (e.g. $\sim$100\%) in the total magnetic field
energy of the star. The SGR 1900+14 observations also strongly argue
against magnetic fields as the source of energy for SGR bursts, since
that should lead to a decrease in the field energy, and hence a
decrease in the spin-down rate during bursting periods. The
observations show the spin-down rate increases.

\section{SGR 1900+14 \& 1806-20 Spin-Down Due to Relativistic Winds}

The association of SGRs 1900+14 and 1806-20 with radio supernova remnants
provides additional evidence as to the origin of their spin-down torques.
Assuming that the spin-down torque is given by $\dot{\Omega} \propto
\Omega^n$, the age of a pulsar with period $P$ and spin-down torque 
$\dot{P}$ is given by

\centerline{$ t_{age} = P/[(n-1)\dot{P}]$}
\noindent where the spin-down braking index, $n$ = 3 for pure magnetic
dipole radiation and $n \sim$ 1 for wind torques.  Taking a nominal age
of 10 kyr for a detectable \cite{bgl89} supernova remnant, inverting
the age equation yields

\centerline{ $n = 1 + (P/\dot{P}) (t_{age})^{-1}$.}
\noindent 
Using $P$ = 5.16 s and the long-term $\dot{P}$ = 6$\times$10$^{-11}$ s/s 
measured for SGR 1900+14, we find that

\centerline{ $n = 1 + 0.27/(t_{age}/10^4$yr),}
and $P$ = 7.47 s and the long-term $\dot{P}$ = 8.3$\times$10$^{-11}$ s/s 
measured \cite{kou98} for SGR 1806-20

\centerline{ $n = 1 + 0.29/(t_{age}/10^4$yr).}
\noindent This indicates that the braking index for SGRs 1900+14 and
1806-20 must be $\sim$1, and that the spin-down of SGR 1900+14 is
dominated by torques due to the relativistic wind and not magnetic
dipole radiation.

\section{Spin-Down Torques of SGRs}

Addressing these problems, Thompson et al \cite{th99} considered a 
magnetar driven Alven wind. The torque provided by the emission of 
such a relativistic wind is 

\centerline{ $I_* \dot{\Omega}_w = -\Lambda(L_w/c^2)R_A ^2 \Omega$}

\noindent where $I_*$ is the neutron star moment of inertia, $L_w$ is
the mechanical luminosity of the wind, $\Omega \equiv 2\pi/P$ is the spin 
frequency, $\dot{\Omega}_w$ is the spin-down rate due to the wind, and 
$R_A$ is the Alfven radius.  $\Lambda$ is a constant equal to 2/3 for a magnetic
dipole field aligned with the rotation axis.
The Alfven radius is given by:

\centerline{ $\frac{L_w}{4 \pi R_A ^2 c} = \frac{B_*^2 (R_A)}{8 \pi}$}
\noindent where $B_*$ is the magnetic field of the neutron star.
When the Alfven radius is inside the light cylinder radius ($R_A < R_{lc}$, where $R_{lc} = c/\Omega$),

\centerline{ $I_* \dot{\Omega}_w = - \Lambda B_* R_*^3 \left( \frac{L_w}{2c^3} \right)^{1/2} \Omega$}
\noindent where $R_*$ is the radius of the neutron star and dipole geometry is assumed.
When the Alfven is outside the light cylinder radius, the torque is limited to

\centerline{$I_* \dot{\Omega}_w = - \Lambda L_w \Omega^{-1}$}
\noindent The transition frequency between these two wind spin-down regimes is

\centerline{{ $\Omega_{tr} = 8.572 \left( \frac{L_w}{10^{36} \rm{ergs/s}} \right)^{1/4} \left( \frac{B_*}{10^{14} \rm{G}} \right)^{-1/2} $} radians/s.}
\noindent The torque due to a rotating magnetic dipole is:

\centerline{ $I_* \dot{\Omega}_{mdr} = - k \frac{B_*^2 R_*^6}{6c^3} \Omega^3$}
\noindent where k = 1 \cite{hck99}.

Once the total spin-down torque is specified as a function of $\Omega$,
the age of the SGR can found by the integral of $d\Omega$ over the total
torque divided by $I_*$, where the integration is performed from an initial
frequency to the present-day angular frequency.

\newpage

\begin{figure}[H]
\centerline{\psfig{file=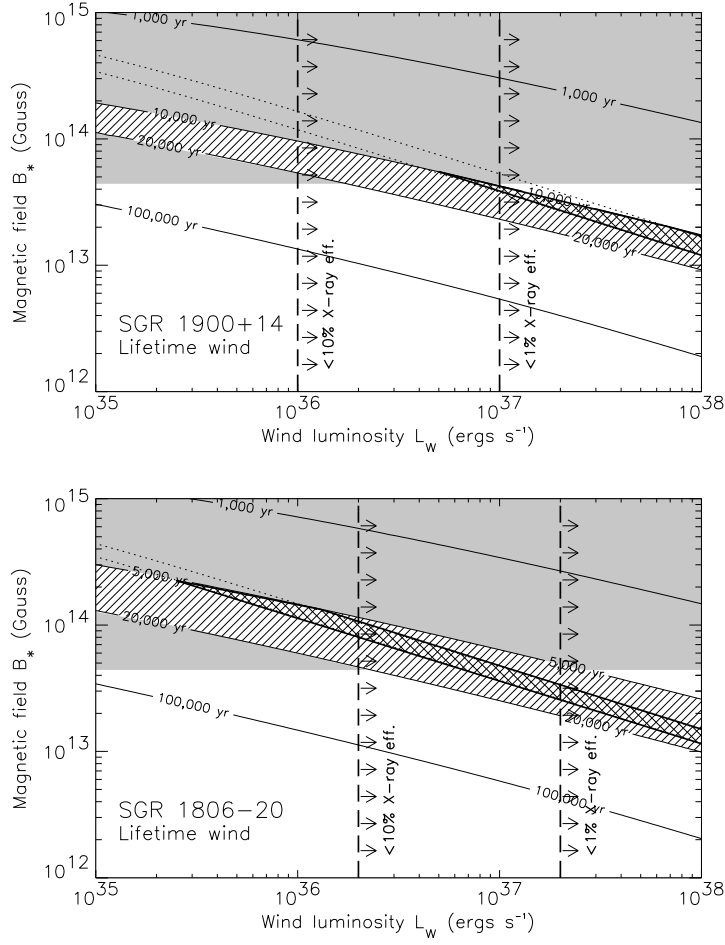, angle=0, width=10cm}}
\caption[]{Age contours for SGRs 1900+14 and 1806-20 for constant magnetic 
fields and Alfven wind luminosities. The cross-hatched areas denote the
allowed regions of parameter space given the constraints provided by
the age of the associated supernova remnant (solid lines) and long term
present-day spin-down rate (dotted lines). The vertical dashed lines
denote the 10\% and 1\% efficiencies of the wind in producing the
observed x-ray flux of $\sim$10$^{35}$ ergs/s for SGR 1900+14 and
$\sim2\times$10$^{35}$ ergs/s for SGR 1806-20.}

\end{figure}

\section{Magnetic Field and Wind Luminosity Limits}

Using the above model we explore a wide range of magnetic fields $B_*$
and wind luminosity $L_w$, shown in Fig. 2. We see in the upper panel
that the presently observed period of $P$ = 5.157 s, the spindown rate
of $\dot{P} = 6\pm1\times10^{-11}$ s/s (dotted lines) of SGR 1900+14,
and the 10 to 20 Kyr range of ages (solid lines) of its associated
supernova remnant G42.8+0.6 \cite{va94}, tightly constrain the
allowable magnetic field to $B_* < 6\times10^{13}$ G and wind
luminosities $L_w > 5\times10^{36}$ erg/s. Compared to the quiescent
2-10 keV x-ray luminosity of $\sim 10^{35}$ erg/s \cite{mur99}, this
wind luminosity implies a $<$ 2 \% conversion efficiency of wind energy
to x-rays in that band which is quite consistent with theoretical
calculations \cite{hck99,ta94,har95}.  We also see in the lower panel 
of Fig. 2, that a similar set of constraints can be obtained for
SGR 1806-20 and its supernova remnant G10.0-0.3, using the present spin 
period \cite{kou98} $P$=7.47 s and $\dot{P}$ = 8.3$\times$10$^{-11}$s/s. 
Again, a wind luminosity with $<$2\% conversion efficiency to x-rays 
yields a sub-critical ($4\times10^{13}$ G) magnetic field. The limit 
for SGR 1806-20 is very similar that found from a comparable analyses
by Harding et al. \cite{hck99}. Thus, we see that with such winds 
the magnetic field limits are quite consistent with the limiting values 
inferred for normal radio pulsars, but not with those expected for magnetars. 

\section{ Summary}

We show that the large variations in the spin down rate $\dot{P}$ 
measured in SGRs and AXPs argue strongly against spin-down by 
magnetic dipole radiation, since that would require large increases 
(e.g. $\sim$100\%) in the total magnetic field energy of the star. 
The SGR 1900+14 observations also strongly argue against magnetic 
fields as the source of energy for SGR bursts, since that should 
lead to a decrease in the field energy, and hence a decrease in the 
spin-down rate during bursting periods rather than the increase observed.
We further show from the braking indicies that the spin-down torque
of the two SGRs is in fact due to winds, not magnetic dipole radiation, 
and that with such the magnetic fields are $<6\times10^{13}$ G,  which
is quite consistent with normal pulsars, but not with magnetars.

\end{document}